\documentclass[12pt]{article}
\usepackage{amsmath,amssymb,pgf,pgfarrows,pgfnodes,subfig,float,appendix, hyperref}
\usepackage[margin=0.7in]{geometry}

\title{{\rm\footnotesize \qquad \qquad \qquad \qquad \qquad \ \qquad \qquad \qquad \ \ \ \ \ \                  UTTG-10-13\ TCC-007-13     RUNHETC-2013-10     
SCIPP 13/06}\vskip.5in     No Firewalls in HST or Matrix Theory}
\author{Tom Banks\\
Department of Physics and SCIPP\\
University of California, Santa Cruz, CA 95064\\
{\it and}\\
Department of Physics and NHETC\\
Rutgers University, Piscataway, NJ 08854\\
E-mail: \href{mailto:banks@scipp.ucsc.edu}{banks@scipp.ucsc.edu}
\\
\\
Willy Fischler\\
Department of Physics and Texas Cosmology Center\\
University of Texas, Austin, TX 78712\\
E-mail: \href{mailto:fischler@physics.utexas.edu}{fischler@physics.utexas.edu}}

\begin{document}
\maketitle

\begin{abstract}
We use the formalisms of Holographic Space-time (HST) and Matrix Theory\cite{bfss} to investigate the claim of \cite{amps} that old black holes contain a firewall, {\it i.e.} an in-falling detector encounters highly excited states at a time much shorter than the light crossing time of the Schwarzschild radius. In both formalisms there is no dramatic change in particle physics inside the horizon until a time of order the Schwarzschild radius. The Matrix Theory formalism has been shown to give rise to an S-matrix, which coincides with effective supergravity for an infinite number of low energy amplitudes.
We conclude that the firewall results from an inappropriate use of quantum effective field theory to describe fine details of localized events near a black hole horizon. In both HST and Matrix Theory, the real quantum gravity Hilbert space in a localized region contains many low energy degrees of freedom that are not captured in QU(antum) E(ffective) F(ield) T(heory) and omits many of the high energy DOF in QUEFT.\end{abstract}

\section{Introduction}

The purpose of this paper is to sharpen our argument that quasi-local holographic models of black holes do not exhibit the firewall phenomenon that the authors of \cite{amps} (AMPS) claim follows from a quantum field theoretic analysis of Hawking radiation in a regime in which quantum field theory can be trusted.  While we have some sympathy with the arguments of \cite{nomura}\cite{verlinde}\cite{haydenharlow}, that the thought experiment envisaged by AMPS cannot, for a variety of reasons, be carried out even in principle, our own view is that the problem lies primarily with the use of QFT to analyze fine grained aspects of the quantum information involved in black hole evaporation.

In \cite{holofirewall} we presented these arguments, but we failed to emphasize sufficiently what the important points were, because our understanding had evolved in the course of writing the paper.  We attempt to repair this here.  In addition, we present a completely different set of arguments based on the Matrix Theory models of black holes\cite{bfssbh} in highly supersymmetric compactifications of M-theory.
The details are somewhat different, but the conclusions are similar.  There is no firewall, though the model has quantum states that share all of the familiar properties of black holes, and is manifestly unitary.  There is also a large body of evidence that Matrix Theory does reproduce the correct scattering matrix for low energy effective supergravity.  

The problem of accounting for black hole entropy is one in which there is 
an evident breakdown of local quantum field theory in the low energy regime.
In the absence of a black hole, local quantum field theory can account for at most $o(A^{3/4})$ (in 4 dimensions) of the entropy allowed by the covariant entropy bound in a causal diamond whose holographic screen has area $A$.
In the presence of a black hole, field theory instead over-counts the entropy, and finds an infinite entropy per unit area, despite the fact that, in the vicinity of the horizon of a large black hole, the local space-time geometry of the hole is identical to that of flat space.  Indeed, the same infinity is found for the entropy encountered by a highly accelerated Rindler observer in flat space.
The infinite entropy per unit area comes from modes of arbitrarily short wavelength in regions that have a small space-like separation from the horizon.   These modes have low energy from the point of view of an accelerated observer.  However, if we consider a small causal diamond surrounding a portion of the horizon (Rindler or Schwarzschild) and insist that the state of QFT in that diamond be such that gravitational back reaction is negligible, then the short wavelength modes must be frozen into the Minkowski ground state.  In that state, there is infinite entanglement entropy per unit area of the holographic screen, between DOF localized in the diamond, and DOF an infinitesimal space-like distance outside it.   There is an obvious paradox here.  For the Rindler or Schwarzschild observer, the entropy refers to real excited degrees of freedom, while entanglement entropy is a property of a pure state.

Even in QUEFT, this paradox is resolved by noting that the Rindler and Schwarzschild observers use a different Hamiltonian from the geodesic observer in a locally Minkowski space. The Rindler case is particularly illuminating because observers with different acceleration see different temperatures.   If we make the assumption that observers sharing the same causal diamond must see the same quantum state up to a unitary transformation, we are prodded in the direction of the conclusion that each accelerated trajectory must have its own Hamiltonian, since they each see a different von Neumann entropy.  This is the starting point for the axioms of HST.  We will not repeat the HST description of accelerated observers here, but refer the reader to \cite{holounruh}. However, we do pause to record the message of that paper: {\it A causal diamond of finite area $\sim N^2$ in a 4-dimensional space-time has $o(N^2)$ (graded)-commuting copies of a super-algebra with a finite dimensional unitary representation. 
The fermionic generators of that algebra are labeled $\psi_i^A (P)$, where the explicit indices are those of an $N \times N+1$ matrix. At most $o(N^{3/2})$ of these DOF have a particle interpretation, and it is only to this subset that the rules of QUEFT apply. We call the rest, {\it horizon DOF}. By contrast, a Jacobsonian\cite{ted} TH(ermodynamic) E(ffective) F(ield) T(heory) will always encode the coarse grained hydrodynamics of any Lorentzian space-time in HST\footnote{Jacobson showed that, apart from the cosmological constant, Einstein's equations follow from the first law of thermodynamics, applied to a system with an effective space-time description, and such that the entropy seen by a maximally accelerated Rindler observer, near each point, varies along the observer's trajectory like the area transverse to a bundle of initially parallel trajectories. HST satisfies Jacobson's criteria, and fixes the c.c. by a boundary condition relating the behavior of area vs. proper time in the limit of a diamond with large proper time.}.  For a geodesic observer in a causal diamond, the bulk of the degrees of freedom have a coupling of order $1/N$ to the particles.  They give rise to particle interactions, but have no long term entanglement with properly prepared incoming particle states.  Accelerated observers have different Hamiltonians, and experience more entanglement: the redshift of particle energies compared to those of horizon DOF couples the particles more strongly and leads, in the large $N$ limit, to the Unruh temperature.}

The common practice in studies of black holes using QUEFT, has been to cut off the infinity of low energy QUEFT DOF in the accelerated observers frame
by putting a ``brick wall" a finite space-like distance away from the horizon.  Then, the remaining finite entropy per unit area is divided between a stretched horizon whose dynamics is admitted to be unknown,  and a ``zone" where standard QUEFT descriptions are valid.  This is the description used by AMPS.  We do not really have a quarrel with this prescription, if it is viewed as a relatively coarse grained model of the black hole.   What we claim is that such a model cannot pretend to account for the quantum dynamics of the tiny fraction ($< o(N^{3/2})$) of the DOF, which are all that is necessary to describe particle physics inside the horizon of the black hole.

\section{Black Holes in HST}

In HST, space-time physics is described in terms of an infinite number of quantum systems, each of which encodes the physics as seen along a particular time-like trajectory, in a proper time dependent Hamiltonian.  Relations between density matrices for shared information, combined with the holographic connection between Hilbert space dimension and area, enable us to extract the causal structure and conformal factor of a Lorentzian metric from the quantum mechanics.  

When we first approached this problem, we believed that the main feature of HST, which avoided the firewall, had to do with the fact that different observers had different Hamiltonians.  J. Polchinski and D. Harlow convinced us that the problem could be recast entirely in the Hilbert space of the observer that is called Alice in most of the firewall literature.  Our resolution of the problem in \cite{holofirewall} in fact relied on different features of HST, but  
enough of our early thinking survived in the final draft of that paper, that it has obscured the issue.  What follows is the description of a decaying black hole in HST, purely from the point of view of the detector $A$, formerly known as Alice.
In fact $A$ belongs to a one parameter family of detectors $A(T)$, parametrized by the amount of proper time, after the formation of the horizon\footnote{In HST, the definition of space-like slices has to do with a convention of synchronization of clocks for space-like separated portions of different time-like trajectories.}, before the detector falls through the horizon of the partially evaporated black hole.  To keep everything finite, we introduce a large time $- N$, prior to horizon formation, and designate the time of horizon formation by $0$.   All of the relevant events occur within a causal diamond of area $N^2$.  The following occurs in real-time:

From the point of view of the A detector, the black hole evaporates for a long time before the detector crosses its ``horizon".  In HST we understand that the black
hole space-time is an approximate, hydrodynamic description of some of the quantum degrees of freedom in the S-matrix theory of asymptotically flat space-time.  It is appropriate for a description of those causal diamonds where the detector whose Hamiltonian we are describing receives signals from the black hole horizon.  These signals cannot come from particle degrees of freedom inside the horizon.  That is the lesson encoded in the classical geometry.  In HST, we view those signals as originating from the $o(N^2)$ DOF on the black hole horizon, but, to the extent that the classical geometry tells us that particles inside the horizon
experience ordinary particle physics, we consider the state of those $< o(N^{3/2})$ particle DOF, to be unentangled with the bulk of the horizon states.  Note that the Hamiltonian of the detector A does not interact with these interior particle DOF until the detector's trajectory crosses the horizon. The $o(N^2) - o(N^{3/2})$ DOF with which it {\it does} interact include both those that are associated with the stretched horizon, and those in the ``zone", in the AMPS description. Cutting off the infinite entropy per unit area in QUEFT leaves a huge hole through which an entropy sub-leading in the area can fit.

QUEFT describes the single vacuum state in a region in a single causal diamond in terms of a huge Hilbert space of ``particle states" viewed by an accelerated observer.  We know that in Minkowski space it is incorrect to view an area's worth of entropy in this Hilbert space as describing actual, physically accessible states of the system.  Yet this is precisely what AMPS do.  Their argument is based on the idea that one can understand the restrictions on the utility of QUEFT in terms of a simple UV cutoff on local wavelengths.  In fact, for a geodesic observer in a low curvature region of space time, this claim is simply false.  Instead, as we have argued above, any entropy in the causal diamond that is proportional to the area, must consist of states which have very low energy according to the Hamiltonian of the geodesic observer, {\it and which therefore cannot manifest as particles in the bulk of the diamond}.  This is true for diamonds inside the horizon, outside the horizon, or straddling the horizon.

The utility of the picture of black hole states as {\it partially described by QUEFT with a brick wall cutoff}, is due to the ease with which one can understand Hawking radiation in this picture, and was of course the description used by Hawking in his original argument.  This often leads to the erroneous claim that if we give up this picture of the Hilbert space, we will no longer understand the thermal nature of black holes and the calculation of the Hawking temperature.  In fact, the thermal nature of Hawking radiation implies that the coarse grained details of Hawking's calculation will be reproduced by {\it any} model of a black hole that exhibits it as an ergodic system\footnote{We use the word ergodic very loosely here, not in its technical mathematical sense.  There are a variety of hypotheses about the nature of a quantum system, which lead to certain aspects of thermal behavior.  Any one of them will do.}, with the correct entropy/energy/size relations. 

In fact, Hawking's derivation of Hawking radiation is not valid even for strongly coupled field theories.  The reason that we {\it know} that the thermodynamic picture of black holes is correct is that the Hartle-Hawking state of a black hole is a thermal state, no matter what the field theory is.  The entropy and temperature of the black hole are encoded in its classical geometry.  We think that the deepest understanding of why this is so comes from Jacobson's observation that, apart from the cosmological constant, Einstein's equations are the hydrodynamics of space-time, assuming that space-time emerges from a quantum system obeying the Covariant Entropy Conjecture that the entropy of the Hilbert space of quantum gravity is one quarter the area in Planck units of the holographic screens of infinitesimal causal diamonds around every point\footnote{Here infinitesimal means much larger than Planck scale but much smaller than the local radius of curvature of space-time.}.  

The Jacobsonian point of view also sheds light on the following vexing question:  On the one hand, a finite entropy black hole can be formed by scattering a finite number of particles in Minkowski space, and on the other hand, it seems to have a distinct space-time metric.  The Jacobsonian interpretation of the black hole metric is as a hydrodynamic approximation to the behavior of a large number of DOF in the Minkowski system, {\it whose behavior is not well modeled as particle physics.}.

Our goal in this section is to provide a statistical mechanics, which is in agreement with the thermodynamic predictions of this classical metric.  For us, these include the behavior of in-falling particle systems before they hit the classical singularity.  This was done in \cite{holofirewall}, so we just summarize it here.  We describe the Hamiltonian of the detector $A(T)$, which follows a trajectory that encounters the instantaneous black hole horizon a proper time $T$ after the original black hole horizon forms.  
The black hole metric is approximated by a time dependent sequence of Schwarzschild metrics with a radius parameter $R(t)$ that follows Hawking's evaporation law.  

At any given time,$t$, we split off $\pi (R(t) M_P)^2$ black hole degrees of freedom from the vastly larger number $N^2$, which describe $A(T)$'s full causal diamond.  For $t < T$, the horizon crossing time of $A(T)$ the bulk of these black hole DOF are given a Hamiltonian with a Planck scale time dependence, which is a sum of traces of the matrices  $(1 - \Pi) \psi \psi^{\dagger}(1 - \Pi )$.  $\Pi$ is a projection matrix on a $K(t) \times K(t)$ subspace with $K \leq (R(t) M_P)^{3/2}$. The rest of the $o(N^2)$ DOF, which are not associated with the black hole,  are given a Hamiltonian, and an initial condition in the remote past that are appropriate for describing particle physics in Minkowski space.   We do not have a complete description of this but a class of Hamiltonians that give the right qualitative physics was described in \cite{holounruh}.  There are additonal constraints necessary to guarantee that in the large $N$ limit, the S-matrix becomes super-Poincare invariant, which we have not yet implemented.  Our model does not try to describe the formation of the black hole from some particular incoming particle state.  Although AMPS assume a black hole formed in this way, they assume nothing about the incoming state besides its purity.  Note that, in addition to the rapid variation of the Hamiltonian mixing up the $ [\sqrt{\pi} R(t) M_P - K(t)]^2 $ states there is a much smaller time dependence of the Hamilton, coming from the time dependence of $R(t)$.  The total Hamiltonian is $$H_{Mink} + H_{Hor\ (t)} + H_{K(t)} ,$$ where the second two terms act on the black hole DOF.  As $R(t)$ varies, $H_{Hor\ (t)}$ and $H_{K(t)}$ act on fewer DOF, and we add those to the Hamiltonian $H_{Mink}$.  This time dependence is an addition to the rapid time dependence of $H_{Hor\ (t)}$, which acts on $o( R(t) - K(t))^2$ DOF.  

The Hamiltonian $H_{K(t)}$ describes the evolution of DOF that will be experienced as particles by the detector $A(T)$.  As in all HST models, the time dependent Hamiltonian of the detector splits into two pieces
$$H(t) = H_{in} (t) + H_{out} (t) .$$ The Hamiltonian $H_{K(t)}$ is part of $H_{out} (t)$ until $t = T$.  For $t > T$ it is included in $H_{in} (t)$.  This is not really a discontinuous transition.  The DOF included in $H_{K(t)}$ are, at early times just particles that have been sent in from past infinity and are initially causally separated from the detector $A(T)$.   If we consider the large causal diamond of a geodesic observer that is causally connected to the particles at early times, but never falls into the black hole, then, at early times, these particles are described by the geodesic Hamiltonian and initial conditions in that diamond\cite{holounruh}.  For simplicity, we assume that  they don't undergo any scattering before they enter the causal diamond where $A(T)$ will encounter them.  

The particles in $H_{K(t)}$ are those with which the detector $A(T)$ can interact, between the time it crosses the instantaneous horizon and the time its trajectory encounters the black hole singularity.  Thus, for times $T < \tau \ll T + R(t)$, the state in the Hilbert space representing the entire black hole is approximately a tensor product of a state acted on by $H_{K(t)}$ and one acted on by $H_{Hor\ (t)}$.  In terms of the matrix DOF $\psi_i^A$, these two Hamiltonians are functions of $\Pi \psi \psi^{\dagger} \Pi$ and $(1 - \Pi ) \psi \psi^{\dagger} (1 - \Pi )$.  Interactions between the two sets of DOF via off block diagonal matrix elements, are suppressed by the large number $1/N$, as long as $\tau \ll R(t)$.

It's important to stress why we're making these claims, especially the last one.  Our aim is to construct a quantum system, whose behavior mimics the classical space-time picture of the interior of the black hole.  In that picture, the particles described by the DOF in $H_{K(t)}$, behave, for a time of order $R(t)$, approximately as they would in flat space.

At the time $T$ (within a tolerance $\ll R(T)$) the DOF in $H_{K(t)} $ are incorporated into $H_{in} (t)$ of the detector $A(T)$.  The Hilbert space of that detector is vast, with entropy of order $N^2$.   However, our model of this detector's behavior is that it's own components, and all the particles with which it interacts {\it after crossing the instantaneous horizon}, are described by the Hamiltonian $H_{K(t)}$.  In addition to the slow time dependence induced by the change of $K(t)$and $R(t)$, this Hamiltonian has a time dependence which becomes extremely rapid as $t$ approaches $T + R(t)$.  These time dependent terms are traces of products of the full $R(t)M_P \times R(t)M_P$ matrix $\psi \psi^{\dagger}$ .  They are very small at $t = T$ and become competitive with the ordinary particle physics contributions in a time of order $R(t)$.  Their effect is to mix up the particle DOF with the horizon, so that the distinction between particles and horizon is no longer meaningful.  At a time of order $T + R(t)$ the detector $A(T)$ 
has ``hit the singularity".  

The bulk of the Hilbert space of $A(T)$ knows nothing about this catastrophe, either before or after it happens.  The DOF in this Hilbert space interact with the horizon DOF in the matrix $(1 - P) \psi \psi^{\dagger} (1 - P)$.   If the infinitely intricate measurements envisioned by AMPS could actually be carried out one would find that the assumption in Page's discussion\cite{Page} of information extraction from a black hole was subtly wrong at the time $T$, when $T$ is of order the Page time.  That is, the black hole is not in a generic state of a Hilbert space of entropy $\pi (R(T)M_P)^2$, because a tiny tensor factor whose entropy is of order $K^2 (T) < (R(T) M_P)^{3/2}$ is not entangled with the rest of the space.  If we synchronize the external clock to the proper time of the in-falling trajectory of $A(T)$, then this factor becomes entangled with the rest of the black hole Hilbert space at a time of order $T + R(T)$.  

Clearly, neither the thermodynamic properties of the black hole, nor the statement that the evaporation process is unitary are violated by this model.  One of the assumptions of Page's argument is modified by an amount that is thermodynamically negligible.  The centerpiece of the AMPS argument, the representation of the Minkowski vacuum state in a local region of space-time as an entangled state in a certain factorized basis of the field theory Hilbert space, simply does not appear in the HST formalism.  Our model clearly treats the black hole as a thermodynamic object with the correct energy-entropy-size relations.  So it's simply untrue that one needs this entangled picture to obtain Hawking radiation.

We believe that the only valid criticism of our model is that we have not yet shown that it really reproduces the results of field theory in conventional situations where no black holes are involved.  We argued that this was the case, to the best of our current ability, in \cite{holounruh}, but we are aware that only a complete calculation of some scattering amplitude will really make the case.  In the next section, we will argue that the Matrix Theory description of 11 dimensional Schwarzschild black holes, gives a picture of black hole evaporation consistent with the one we proposed in HST.  There is no firewall, no description of the Minkowski vacuum as an entangled state, a manifestly unitary S-matrix for particle scattering, and manifest super-Galilean invariance.  In addition, it has been shown that an infinite number of scattering amplitudes in this model coincide with those given by a super-Poincare invariant QUEFT - 11 dimensional supergravity.

In principle, the above criticism might be applied to our claims about reproducing the properties of Hawking radiation.  However, we demonstrated in \cite{holounruh} that the Minkowski Hamiltonian, which acts on $o([N - R(t)]^2)$ DOF, acts, as $N \rightarrow\infty$, like the kinetic term of a collection of massless particles, on a tensor factor of entropy $\leq N^{3/2}$ of its Hilbert space.  We also argued that in the large $N$ limit, pure states of these particles remain pure and that the effect of interaction with the bulk of the $o(N^2)$ DOF could be encoded in particle interactions that were localized in space-time.  Thus, our model does describe  black hole evaporation as a sequence of quasi-equilibrium states of the black hole, interacting with a gas of relativistic particles of (potentially) much higher entropy.  The dynamics is explicitly rotation invariant and there are emergent super-Poincare generators, which act on particle states in the large $N$ limit.  This is enough to establish the thermal nature of the spectrum of evaporated particles, even though we have not established that the S-matrix of those particles is Poincare invariant.  Thus, we certainly cannot claim that our model will reproduce the gray body factors that arise in the field theory treatment of Hawking radiation, but the thermal nature of the spectrum and the correct temperature are guaranteed.

In HST itself, we still have to discuss the consistency conditions between detectors $A(T)$ with different values of $T$.  In \cite{holofirewall} we showed that consistency between detectors with $T \sim R_S$ and $T \gg R_S$ implied that the late falling detector had to encounter a singularity\footnote{Quantum translation: the time dependent Hamiltonian must mix the particle DOF inside $A(T)$'s horizon with the horizon DOF in a time of order $R_S$.} in a time of order $R_S$ after it crosses the horizon.  Consistency of the in-falling detector's description with that of a supported detector implies that the description of the black hole from the supported 
detector's point of view must include a tensor factor of entropy $ \sim (R(t)M_P)^{3/2}$, which is approximately unentangled with the horizon\footnote{We thank D. Harlow for repeatedly emphasizing this point to us.}.  Since $(RM_P)^{3/2} \ll (RM_P)^2 $, this does not affect the thermodynamics of black holes until they are of Planck size.  Since $R(t)$ goes to zero eventually, there is no problem with unitarity of the S-matrix either.  Of course, once the black hole is Planck size, the approximate descriptions in this paper lose their validity.  Indeed, even in Minkowski space, the clean separation between particle and horizon DOF is impossible in small causal diamonds.  Indeed, as shown in \cite{holounruh}, this feature of the HST description is responsible for reactions that change the number of particles, and their momenta.  Particles are emergent phenomena in HST, strictly speaking defined only in the limit of infinite causal diamonds.  The validity of QUEFT, with its implicit assumption of infinite numbers of particle states, is even more restricted.

\section{Black Holes in Matrix Theory}

We will restrict attention to the Matrix Theory for String/M-theory in 11 non-compact dimensions.  This is the quantum mechanics of the zero modes of maximally supersymmetric $SU(N)$ Yang-Mills theory.  Similar results would be obtained for compactification of the theory on tori of dimension $1-3$.
The Lorentz invariant limit is achieved by taking $N\rightarrow\infty$ and computing the S-matrix for states whose energy scales like $1/N$.   However, in \cite{bfssbh}, the authors argued that one could understand the qualitative dynamics of black holes of entropy $S$ in the model with $N\sim S \gg 1$.

The construction of black hole states begins with a classical solution of the matrix equations $X_{cl} (t)$, whose variation away from the origin of transverse coordinates was bounded by a distance in Planck units of order one (in the sense of large $N$ counting).  The matrices in the solution have rank $N$. For large $N$, we can think about them using the correspondence \cite{DHN} with light front membrane theory.  The background defines a (fuzzy) toroidal membrane, whose volume is parametrized by two angles $p,q$. The matrices are functions on the phase space $[p,q]$ with commutator given, in the large $N$ approximation, by Poisson brackets. There are many such solutions, but in $11$ dimensions, this multiplicity gives rise to a sub-leading correction to black hole entropy.

Now write $X = X_{cl} + \sum_i  x_i M_i$, where
$$ M_i  = \sum_{k,l} e^{- N[ (p - p_i - 2\pi k)^2 + (q - q_i - 2\pi l)^2]}.$$
The commutators of these matrices satisfy
$$ |[M_i , M_j]| \leq e^{ - \frac{1}{2}[ (p_i - p_j)^2 + (q_i - q_j)^2 ]} .$$  Taking a distribution of points separated by distances of order $\frac{1}{\sqrt{N}}$ (there are $o(N)$ such points on the torus) , we can make these commutators as small as we like.  Thus, there's a basis in which all the matrices $M_i$ are simultaneously block diagonal. The traces of these matrices are $o(1)$.

The commutator between the classical background $X_{cl} (p,q)$ and the fluctuations $ x_i$ gives rise to a harmonic potential binding the $x_i$ to the background.  The terms bilinear in different $x_i$ are of the same order as the commutator $[M_i , M_j ]$ and we drop them. The quadratic potential is
$$\sum_i x_i^2 N \int\ dp\ dq\ [p^2 + q^2] (\nabla X_{cl})^2 e^{ - N (p^2 + q^2)}.$$ The factor of $N$ in front of the integral is the combination of a $1/N$ in the translation of traces of matrix commutators into integrals of Poisson brackets over the membrane, and two factors of $N$ coming from converting derivatives of $M_i$ into factors of $p$ or $q$.  For a smooth classical membrane configuration, the gradient of $X_{cl}$ is $N$ independent for large $N$.  The harmonic potential thus has an overall coefficient $1/N$.  This is, it will turn out, negligible compared to other contributions to the energy.

The latter come from integrating out off diagonal matrices between the different $x_i$ terms.   If the $x_i$ velocities are small, which is verified self consistently, the effective Hamiltonian is

$$H = \sum_i {\bf p_i}^2 + A G_N \sum_{i,j} \frac{\bf (p_i - p_j)^4}{|{\bf x_i - X_j}|^7} .$$  The coefficient $A$ is of order $1$.  For a bound system with this Hamiltonian, and large $N$, the mean interparticle distance $R_S$, the energy per particle, and the total light front energy (which gives us the mass) may be calculated crudely by using the uncertainty principle and the virial theorem.  The result is
$$B G_N^{-1} R_S^9 = N,$$
$$E_{per\ particle} \sim R_S^{-2},$$
$$M \sim G_N^{-\frac{1}{9}} N^{\frac{8}{9}} .$$ Here $R_S$ is the average separation, which is also the size of the bound state.
These are the expected relations for an $11$ dimensional Schwarzschild black hole, with the individual $D0-$branes having the kinematics expected for Hawking particles boosted to the light front frame, and $R_S$ the Schwarzschild radius, if $N$ is indeed the entropy of the black hole.

The fact that $N$ is indeed the entropy follows from the fact that the $D0$ branes are {\it tethered} to different positions on the classical membrane, so that they are in fact distinguishable particles, obeying Boltzmann statistics.  If we plug $x_i \sim R_S$ into the harmonic potential, we find a negligible correction to the total energy.  In the second  paper in\cite{bfssbh}, we showed also that the correct Newtonian interaction between black holes is obtained, if we are careful to note that we are calculating energies averaged over the longitudinal circle of Discrete Light Cone Quantization. In the third paper we estimated the rate of Hawking evaporation, and found agreement with the expectations for a thermal system with the indicated energy and entropy.

The mechanism of Hawking radiation was ``snapping of the tethers": a quantum fluctuation, which momentarily sets to zero the piece of the classical configuration that provides the harmonic binding for a particular $D0-$brane coordinate $x_i$.  That particle then flies out to infinity along the flat direction in the matrix potential.  The black hole then re-equilibrates with one constituent fewer.  

In the second paper of \cite{bfssbh} it was pointed out that the analysis of the dynamics of the non-compact dimensions gave analogous results for black holes in all dimensions.  However, at that time Matrix Theory technology required one to use a more and more complicated field theory to
compactify the theory on more and more dimensions. For a $5$ dimensional toroidal compactification one was forced to go beyond field theory, and for $6$ and more dimensions the Matrix Theory proposal failed.  Results could be established firmly only for compactifications on tori of dimensions $1 - 3$, and in the last of these cases the internal field theory contributed a finite fraction of the black hole entropy.  Recently TB and Kehayias\cite{tbkehayias} have suggested an alternate definition of Matrix Theory, which is a simple quantum mechanics (not a field theory) for every compactification.
It would be interesting to return to the black hole problem using this technology.  We conjecture that a unified qualitative picture of Schwarzschild black hole dynamics might result.

Be that as it may, the purpose of the present paper is to establish the absence of firewalls for 11 dimensional Schwarzschild black holes.  Recall that Matrix Theory defines a scattering matrix for asymptotic states along the $U(k_1) \otimes \ldots \otimes U(k_n)$ flat directions of the Matrix Theory potential.  The $SU(k_i)$ degrees of freedom are frozen into their unique BPS bound state, and the scattering states are manifestly those of eleven dimensional supergravitons.  If we take $k_i$ to infinity, at fixed ratios $\frac{k_i}{k_j}$ then the asymptotic states support an action of the $SO(1,10)$ super-Poincare group.  The S-matrix is manifestly invariant under the super-Galilean sub-group of this group, and the existence of the S-matrix in the limit is equivalent to longitudinal boost invariance.  It is hard to see what kind of instability could make the manifestly unitary S-matrix fail to exist in this limit, because from the point of view of the quantum mechanics, it is a {\it low} energy limit.  In particular, emission of states with longitudinal momentum that does not scale to infinity is forbidden by energy conservation.  

We also know that the S-matrix obeys an infinite number of non-renormalization theorems\cite{dineetal}, which imply that an infinite number of terms in the low energy expansion of the hypothetical limit, actually coincide with those expected from the low energy expansion of a super-Poincare invariant effective Lagrangian.  This proves that the limiting S-matrix is not the unit matrix, and strongly suggests the existence and super-Poincare invariance of all matrix elements.  

Finally, Matrix Theory provides a definition of finite time transition amplitudes, for processes that take place over a finite range of transverse distance.  These amplitudes manifestly approach the corresponding S-matrix elements as the time and transverse distance go to infinity.  Using these, we can model the experience of an apparatus falling into a black hole.   The apparatus is modeled by a $K \times K$ block of the Matrix Theory variables, with $1 \ll K \ll N$.  Initially, we take the transverse separation between the $K \times K$ block and the $N \times N$ block, which represents the black hole, to be very large, and set the center of mass of the $K \times K$ block moving slowly towards the transverse position of the hole.  Consider an initial condition for the $SU(K)$ variables which consists of two groups $K_{1,2}$ of supergravitons coming in from a large distance.  For comparison, we take $K_1 = K_2$. The first group collides at a time long before the c.m. of the block approaches the position of the black hole.  What this means is that the incoming coefficients of a block diagonal matrix of block sizes $p_1 \ldots p_m$, with $\sum p_i = K_1$ become close enough so that it no longer takes a huge energy to excite the off diagonal matrices.   Non-Abelian dynamics becomes important and we find finite amplitudes for going off in flat directions $q_1 \ldots q_n$, with $\sum q_i = K_1 $.  We compute the amplitude from the time the initial separation is $L \ll R_S$ until a time after collision when final separations are of order $L$.  During this entire period of time, the separation between the c.m. of the $K_1 \times K_1$ block and the black holes is $\gg R_S$.

For the $K_2$ block, we instead time the incoming particles so that they begin to interact strongly with each other when the c.m. of the block is within $R_S$ of the center of the black hole.  Since $L \ll R_S$, the typical distance between particles in the block is much less than their distance from any of the $D0-$brane constituents of the black hole.  Thus, the interactions with the black hole can be considered a small perturbation of the particle interactions in flat space, over times where the transverse separation is $\ll R_S$.  Over longer time scales, this is no longer true.  For distance scales of order $R_S$ from the $K_2$ block, a particle we originally considered part of the $K_2$ block will suffer multiple scatterings with black hole constituents, whose transverse separation from it are smaller than or of the same order as
those in the $K_2$ block.  Since the number of black hole constituents is large compared to the number of particles in the original event (since $N \gg K_2$) it is plausible that the particle will come into equilibrium with the black hole constituents.  That is, interactions with the constituents will tend to break it up into its individual $D0-$branes and these will equilibrate and become indistinguishable, in a coarse grained way, from constituent $D0-$branes that were in the black hole before the $K_2$ block approached it.  There is probably a theorem to be proven here in the $N\rightarrow\infty$ limit, since in that limit, the constituents of the $K_2$ block can never escape from the 
black hole once they have come within a distance of order $R_S$ of it.  

We claim that this is evidence for the absence of a firewall in Matrix Theory.  Particle physics over time scales smaller than the time to traverse a transverse distance $R_S$ is affected only perturbatively by the question of whether it takes place inside or very far from the Schwarzschild radius of a black hole.  Note also that {\it nowhere in the Matrix Theory model of a black hole does there exist any analog of the high energy particles that are supposed to constitute the firewall.}  Black hole constituents in Matrix Theory have the kinematic properties of typical thermal Hawking particles.  It is probably significant that it's a general property of physics in a light front frame, that the particle theory vacuum is trivial.  This means that the vacuum entanglement that is claimed to be a crucial feature of Hawking radiation by AMPS cannot be a feature of physics formulated on the light front, as Matrix Theory is.

The reason that this demonstration of the absence of firewalls adds to the credibility of our HST argument is that there is much more evidence that Matrix Theory is a systematic approximation to a super-Poincare invariant S-matrix theory of particles, with a low energy effective field theory expansion.
Furthermore, there is an established construction of states with the properties of black holes, in a system with a time independent Hamiltonian.
The evidence that HST leads to super-Poincare invariant scattering was presented in \cite{holounruh}, and is much less extensive.

\section{Conclusions}

Both of our models of quantum gravity contain ``low energy" DOF, which are not captured by QUEFT, and are crucial to the description of the local dynamics of black holes.  In neither of them is there any apparent hint of the picture of field theory with a stretched horizon cutoff, which pervades much of the literature on the black hole information problem, including the paper of AMPS.

It is not our intention here to claim that the stretched horizon picture is completely wrong or useless.  Rather, our position is that the question of whether an in-falling observer encounters large deviations from flat space physics on a time scale much shorter than the classical in-fall time to the singularity, involves only a tiny fraction of the DOF of the black hole.  QUEFT with a stretched horizon cutoff is certainly a grossly thermodynamic description of the system, and is simply insensitive to these thermodynamically negligible DOF.  To make the AMPS argument one must assume that the stretched horizon QUEFT description is an accurate accounting of the dynamics at the level of single bits.

\vskip.3in
\begin{center}
{\bf Acknowledgments }
\end{center}
T.B. would like to acknowledge conversations with J. Polchinski, L.Susskind and D.Harlow about firewalls.  He would also like to thank the organizers of the firewall workshop at CERN in March 2013, for the invitation to speak at that conference, and the other participants for stimulating discussions. The work of T.B. was supported in part by the Department of Energy.   The work of W.F. was supported in part by the TCC and by the NSF under Grant PHY-0969020


\begin{thebibliography}{99}

\bibitem{amps}A. Almheiri, D. Marolf, J. Polchinski and J. Sully, Black Holes: Complementarity or
Firewalls?, 1207.3123; A.~Almheiri, D.~Marolf, J.~Polchinski, D.~Stanford and J.~Sully,
  ``An Apologia for Firewalls,''
  arXiv:1304.6483 [hep-th].
\bibitem{haydenharlow} D.~Harlow and P.~Hayden,
  ``Quantum Computation vs. Firewalls,''
  arXiv:1301.4504 [hep-th].
  
\bibitem{nomura}  Y.~Nomura, J.~Varela and S.~J.~Weinberg,
  ``Low Energy Description of Quantum Gravity and Complementarity,''
  arXiv:1304.0448 [hep-th].
\bibitem{verlinde}  E.~Verlinde and H.~Verlinde,
  ``Black Hole Entanglement and Quantum Error Correction,''
  arXiv:1211.6913 [hep-th].
  \bibitem{ted} T.~Jacobson,
  ``Thermodynamics of space-time: The Einstein equation of state,''
  Phys.\ Rev.\ Lett.\  {\bf 75}, 1260 (1995)
  [gr-qc/9504004].
  \bibitem{holofirewall} T.~Banks and W.~Fischler,
  ``Holographic Space-Time Does Not Predict Firewalls,''
  arXiv:1208.4757 [hep-th].
  \bibitem{tbkehayias} T.~Banks and J.~Kehayias,
  ``Fuzzy Geometry via the Spinor Bundle, with Applications to Holographic Space-time and Matrix Theory,''
  Phys.\ Rev.\ D {\bf 84}, 086008 (2011)
  [arXiv:1106.1179 [hep-th]].
 \bibitem{bfssbh} T.~Banks, W.~Fischler, I.~R.~Klebanov and L.~Susskind,
  ``Schwarzschild black holes from matrix theory,''
  Phys.\ Rev.\ Lett.\  {\bf 80}, 226 (1998)
  [hep-th/9709091].
  ; T.~Banks, W.~Fischler, I.~R.~Klebanov and L.~Susskind,
  ``Schwarzschild black holes in matrix theory. 2.,''
  JHEP {\bf 9801}, 008 (1998)
  [hep-th/9711005].
  ;  T.~Banks, W.~Fischler and I.~R.~Klebanov,
  ``Evaporation of Schwarzschild black holes in matrix theory,''
  Phys.\ Lett.\ B {\bf 423}, 54 (1998)
  [hep-th/9712236].
  \bibitem{DHN} B.~de Wit, J.~Hoppe and H.~Nicolai,
  ``On the Quantum Mechanics of Supermembranes,''
  Nucl.\ Phys.\ B {\bf 305}, 545 (1988).
\bibitem{Page} D. N. Page, ``Information in black hole radiation," Phys. Rev. Lett. 71, 3743 (1993)
[hep-th/9306083].
\bibitem{bfss}  T.~Banks, W.~Fischler, S.~H.~Shenker and L.~Susskind,
  ``M theory as a matrix model: A Conjecture,''
  Phys.\ Rev.\ D {\bf 55}, 5112 (1997)
  [hep-th/9610043].
\bibitem{holounruh} T.~Banks, W.~Fischler, {\it Holographic Space-time, the Unruh Effect and the S-matrix}, Journal of Emergent Physics {\it to Emerge}.
\bibitem{dineetal}  M.~Dine, R.~Echols and J.~P.~Gray,
  ``Tree level supergravity and the matrix model,''
  Nucl.\ Phys.\ B {\bf 564}, 225 (2000)
  [hep-th/9810021].
; M.~Dine, R.~Echols and J.~P.~Gray,
  ``Renormalization of higher derivative operators in the matrix model,''
  Phys.\ Lett.\ B {\bf 444}, 103 (1998)
  [hep-th/9805007].
  ;  K.~Becker and M.~Becker,
  ``On graviton scattering amplitudes in M theory,''
  Phys.\ Rev.\ D {\bf 57}, 6464 (1998)
  [hep-th/9712238].
  ;  J.~A.~Harvey,
  ``Spin dependence of D0-brane interactions,''
  Nucl.\ Phys.\ Proc.\ Suppl.\  {\bf 68}, 113 (1998)
  [hep-th/9706039].
\end{thebibliography}
\end{document}